\documentclass[%
 reprint,
 amsmath,amssymb,
 aps,
]{revtex4-2}
\usepackage{comment}
\usepackage{placeins}
\usepackage{graphicx}
\usepackage{dcolumn}
\usepackage{bm}

\def\L{\mathcal{L}}
\def\Ca{^{40}\text{Ca}^+}
\def\n{\bar{n}}
\def\nss{\bar{n}_{ss}}
\def\S{S_{1/2}}
\def\P{P_{1/2}}
\def\D{D_{3/2}}
\newcommand{\bra}[1]{\langle #1|}
\newcommand{\ket}[1]{|#1\rangle}

\newcommand{\abs}[1]{ \left\lvert#1\right\rvert}
\begin{document}
\preprint{APS/123-QED}

\title{Trapped-ion Laser Cooling in Structured Light Fields}

\author{Zhenzhong (Jack) Xing}
\email{zx296@cornell.edu}
\author{Karan K. Mehta}%
\affiliation{%
 School of Electrical and Computer Engineering, Cornell University, Ithaca, NY 14853, USA
}%

\date{\today}

\begin{abstract}
Laser cooling is fundamental to quantum computation and metrology with trapped ions. In particular, quantum ground state cooling required for many gate implementations can occupy a majority of runtime in current systems. A key limitation to cooling arises from unwanted carrier excitation, which in typically used running wave (RW) fields invariably accompanies the sideband excitations effecting cooling. Laser cooling in structured light profiles enables selective sideband excitation with nulled carrier drive; motivated by integrated photonic approaches' passive phase and amplitude stability, we propose simple configurations realizable with waveguide addressing using either standing wave (SW) or first-order Hermite-Gauss modes. We quantify performance of Doppler cooling from beyond the Lamb-Dicke (LD) regime, and ground-state cooling using electromagnetically induced transparency (EIT) leveraging these field profiles. Carrier-free EIT offers significant benefits in cooling rate and motional frequency bandwidth, and final phonon numbers comparable to those achievable with resolved sideband cooling. Carrier-free Doppler cooling's advantage is significantly compromised beyond the LD regime but continues to hold, indicating such configurations are applicable for highly excited ions. Our simulations focus on level structure relevant to $\Ca$, though the carrier-free configurations can be generally applied to other species. We also quantify performance limitations due to polarization and modal impurities relevant to experimental implementation. Our results indicate potential for simple structured light profiles to alleviate bottlenecks in laser cooling, and for scalable photonic devices to improve basic operation quality in trapped-ion systems. 
\end{abstract}

\maketitle

Trapped ions constitute a leading platform for quantum computation \cite{bruzewicz2019trapped} and simulation, with recent systems demonstrating algorithms on tens of qubits with arbitrary connectivity \cite{moses2023race, postler2024demonstration}, leveraging the high-fidelity entangling gates \cite{Ballance_Harty_Linke_Sepiol_Lucas_2016, gaebler2016high, clark2021high, loschnauer2024scalable}, state preparation, measurement, and readout \cite{harty2014high} achievable in this platform. The high coherence and precise control of systematics achievable in ion traps also enables timekeeping at the present limits of accuracy \cite{brewer2019al+, Hausser_2025}. 

High-fidelity trapped-ion control in quantum computing systems typically requires laser cooling to near the motional ground state \cite{leibfried2003quantum}, also of interest for next-generation clocks \cite{kulosa2023systematic}. To this end, Doppler cooling may be followed by electromagnetically induced transparency (EIT) cooling \cite{morigi2000ground, roos2000experimental, Eschner_Morigi_Schmidt-Kaler_Blatt_2003}, polarization gradient cooling \cite{Birkl_Yeazell_Rückerl_Walther_1994}, and/or sideband cooling \cite{Diedrich_Bergquist_Itano_Wineland_1989, Srinivas_Burd_Sutherland_Wilson_Wineland_Leibfried_Allcock_Slichter_2019}, to successively bring expectation phonon number $\n$ to near zero \cite{leibfried2003quantum}. Due to inevitable heating of atoms during operation \cite{de2021materials} and because readout and transport operations repeated in trapped-ion quantum computing operation \cite{kielpinski2002architecture} can excite ion motion, ground-state cooling generally must be repeatedly applied and can occupy a majority of runtime in current trapped-ion quantum computing systems  \cite{Pino_Dreiling_Figgatt_Gaebler_Moses_Allman_Baldwin_Foss-Feig_Hayes_Mayer_etal._2021}. Since operation times are a key limitation of trapped-ion systems, cooling schemes with improved cooling rate and final phonon number may address a crucial bottleneck in trapped-ion systems.

\begin{figure*}[t!]
    \centering
\includegraphics[width=0.85\textwidth]{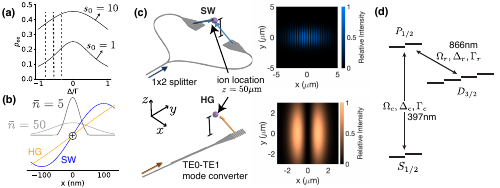}
    \caption{(a) Steady state excitation population of a two-level system as a function of detuning $\Delta$. Carrier $\ket{g,n}\rightarrow \ket{e,n}$, blue sideband $\ket{g,n}\rightarrow \ket{e,n+1}$, and red sideband $\ket{g,n}\rightarrow \ket{e,n-1}$ excitations correspond to black, blue, and red dotted lines respectively. Due to power broadening, higher beam power ($s_0 = 2\Omega_0^2/\Gamma^2$) reduces the contrast between the red sideband and blue sideband scattering rates, compromising cooling performance. (b) Probability distribution of a $\Ca$ ion in thermal states of different $\bar n$ for $\omega_m = 2 \pi \times 3$~MHz, centered at the nodal point of a $\lambda = 397$~nm SW and a $w_0 = 500$~nm TE$_{10}$ Hermite-Gaussian (HG) mode. For larger $\n$, the wave function samples a larger field region, suggesting different cooling behaviors when outside/inside the LD regime. (c) Two experimental schemes for carrier-free driving using integrated photonics, based on two counter-propagating beams from the same input to form a passively phase-stable SW, or a TE$_{10}$ mode delivered from a single grating emitter fed by a TE1 guided mode \cite{Beck_Home_Mehta_2024}. The cooling beam's intensity profiles are sketched on the right. (d) The eight-level structure relevant to $\Ca$ ($\S,\P,\D$) with lasers for Doppler cooling. $\Delta_c, \Delta_r$ are detuning between $\S$ and $\P$, $\D$ and $\P$ respectively at zero magnetic field.}
    \label{fig:Dopplerscheme}
\end{figure*}

Key limitations to cooling performance arise from couplings that accompany the desired interactions when employing the usual plane-wave-like running wave (RW) fields. In a frame rotating with laser frequency $\omega$, interaction between a RW laser field and an atom represented as a two-level system with ground and excited states $\ket g$ and $\ket e$, oscillating in a single motional mode at angular frequency $\omega_m$, can be described with a dipole interaction Hamiltonian expanded as: 
\begin{equation}
\langle e| \hat V_\mathrm{dip} |g\rangle =  \frac{\Omega_0}{2} e^{i\bm{k} \cdot \bm{\hat{R}}} =\frac{\Omega_0}{2} \left(1+i\eta (\hat{a}+\hat{a}^\dag) +... \right)
\label{eq:one},
\end{equation}
for wave vector $\bm{k}$ with $\abs{\bm{k}} = k_0$ in free space; Rabi frequency $\Omega_0$; position operator $\bm{\hat{R}} = \bm{u_x} x_0 (\hat{a}+ \hat{a}^\dag)$ written in terms of the ground-state wavefunction extent $x_0=\sqrt{\frac{\hbar}{2m\omega_m}}$, the annihilation and creation operators $\hat{a}$ and $\hat{a}^\dag$, and the unit vector $\bm{u_x}$; and Lamb-Dicke (LD) parameter $\eta = \eta_0 \cos(\theta)$, where $\eta_0 = k_0 x_0$ and $\theta$ is the angle between $\bm{k}$ and $\bm{u_x}$. The terms in Eq.~(\ref{eq:one}) proportional to $\eta \hat a$ and $\eta \hat a^\dagger$ describe coherent coupling to red and blue motional sidebands, respectively \cite{leibfried2003quantum}, driven by first-order field gradients desired for cooling; larger scattering rate on the red as compared to the blue sideband leads to cooling (Fig.~\ref{fig:Dopplerscheme}a) \cite{Stenholm_1986, Lindberg_Stenholm_1984}. The leading zeroth order term, due to the field at the ion location, represents carrier coupling that does not affect the motional state. Since typically $\eta < 1$, the carrier drive strength exceeds that of the sidebands and is problematic for two reasons. As Rabi frequency is increased to cool more rapidly, this carrier excitation saturates the internal population, reducing contrast in the excitation rate between the red and blue sideband, thereby limiting cooling rates and compromising the final phonon number achievable (Fig.~\ref{fig:Dopplerscheme}a). Additionally, while carrier excitation is associated with no motional excitation, the subsequent spontaneous decay results in net heating. Thus, carrier excitation limits achievable cooling rate and additionally causes net heating. 

These limitations can be alleviated via suppression of carrier excitation and selective sideband driving achievable at appropriate locations in structured light fields \cite{Cirac_Blatt_Zoller_Phillips_1992, Zhang_Wu_Chen_2012, deLaubenfels_Burkhardt_Vittorini_Merrill_Brown_Amini_2015, vasquez_control_2023}. For electric dipole transitions of atoms trapped at points with zero intensity and nonzero field gradient, e.g. at SW nodes or nodal lines of higher-order Hermite-Gaussian (HG) modes, the carrier term along with all even orders of the expansion in Eq.~\ref{eq:one} is nulled. Fig.~\ref{fig:Dopplerscheme}b gives an example with $\Ca$ and its $397{}$ nm dipole transition. The first-order sideband term then becomes the leading contribution in the Hamiltonian, i.e.
\begin{equation}
\langle e| \hat V_\mathrm{dip} |g\rangle = \frac{\Omega_0}{2} \sin(\bm{k} \cdot \bm{\hat{R}})= \frac{\Omega_0}{2}\left(\eta (\hat{a}+\hat{a}^\dag) +...\right)
\label{eq:carrier-free}.
\end{equation}
Note that with this definition the total laser power in the SW configuration is half that of the RW configuration for the same $\Omega_0$.

Recent development of integrated waveguide photonics for optical control in planar ion traps \cite{mehta2016integrated,  niffenegger2020integrated, Mehta_Zhang_Malinowski_Nguyen_Stadler_Home_2020, vasquez_control_2023, ivory2021integrated, kwon2024multi, mordini2024multi} offers a route to realize these mechanisms without the phase drifts and pointing instability that pose significant challenges in conventional free-space implementations \cite{schmiegelow2016phase}. For example, SW generation straightforwardly leverages counterpropagating beams derived from a single waveguide input (Fig.~\ref{fig:Dopplerscheme}c, top) as demonstrated in \cite{vasquez_control_2023}. A TE$_{10}$ HG  mode with transverse field gradients at the intensity node along the beam center (Fig.~\ref{fig:Dopplerscheme}c, bottom) can be achieved via emission of higher-order waveguide modes \cite{Beck_Home_Mehta_2024}. While the form of the interaction Hamiltonian is the same to lowest order for these two cases, the effective LD parameter in the latter case, $\eta_{\text{HG}} = \frac{2\sqrt{2}}{w_0} x_0$ \cite{west2021tunable} reflects motional coupling from transverse field gradients, additionally offering avenues to engineer interactions to accommodate different ion crystal configurations and motional mode orientations. Note that structured light profiles discussed here can also be used to improve trapped-ion operation beyond laser cooling \cite{Saner_Băzăvan_Minder_Drmota_Webb_Araneda_Srinivas_Lucas_Ballance_2023}. While we focus on these relatively simple field configurations, similar considerations would apply to more general structured light profiles \cite{verde2023trapped}.

In this work, we examine the practical performance of Doppler and EIT cooling using structured light fields through quantum optical master equation simulations. We consider $\Ca$'s relevant eight-level structure ($\S,\P,\D$) in Fig.~\ref{fig:Dopplerscheme}d, although the methods analyzed are applicable to other species. Our treatment extends previous theoretical work in considering Doppler cooling from well beyond the LD regime (Fig.~\ref{fig:Dopplerscheme}b), as well as accounting for sideband saturation effects at high drive strengths for both Doppler and EIT cooling. 

In the results presented below we show significant enhancement possible for carrier-free EIT ground-state cooling, simultaneously in cooling rate, motional mode frequency bandwidth, and steady-state phonon number achievable, with over an order of magnitude improvement predicted for the latter. We quantify the sensitivity of these enhancements over typical RW schemes to likely experimental limitations such as polarization and mode-delivery impurities. Realization of these simultaneous advantages may significantly reduce the sideband cooling required to achieve ground-state occupancies desired for high-fidelity optical control. For Doppler cooling, despite the expected degradation in SW cooling of highly excited ions as compared to within the LD regime, we still find that a modest enhancement for SW cooling obtains. While sub-Doppler cooling via polarization gradient cooling can also leverage integrated photonic beam delivery \cite{hattori2022integrated} and benefit from ion positioning at polarization SW nodes, the expected enhancement in steady-state phonon number compared to with RWs is limited to $2\times$ \cite{joshi2020polarization}. Given the more significant gains predicted for EIT here, we focus on EIT as a sub-Doppler and ground-state cooling method. We present our results for the simpler Doppler cooling first followed by EIT cooling, commenting on the impact of $\eta_0$ on both schemes. We demonstrate that carrier-free Doppler and EIT cooling use half of the time of RW Doppler and EIT cooling to sequentially cool the ion from outside the LD regime to near ground state with $\n< 10^{-2}$, without further resolved sideband cooling that usually takes multiple milliseconds. This work indicates potential for scalable photonic hardware to alleviate bottlenecks in trapped-ion control, focusing here on laser cooling, and informs experimental implementation in integrated devices. 

\section{Carrier-free Doppler cooling}
The treatment of \cite{Cirac_Blatt_Zoller_Phillips_1992} showed that Doppler cooling for ions within the LD regime at SW nodes allows higher cooling rates via use of higher cooling beam intensities, as well as lower steady-state phonon number $\nss \equiv \lim_{t\rightarrow \infty}\n$, than possible with RWs. In a RW configuration and in the LD regime, saturation sets in when $\Omega_0 \sim \Gamma$. At a SW null, since only sideband couplings are present, saturation instead occurs when $\eta\sqrt{n} \Omega_0 \sim \Gamma$. Hence, the cooling rate $W_c \equiv -\dot{\bar n}/\bar n\propto (\eta \Omega_0)^2$ can be increased by of order ${\sim1}/\eta^2 n $ as compared to with RW fields. 
This delayed saturation enables a significant gain in $W_c$ for sufficiently low phonon numbers $n$ given that typically,  $\eta\sim 0.1$. Furthermore, an approximately $2\times$ reduction in final phonon number $\nss$ is predicted. 

However, drastically different behavior, including heating, appears at other positions in the SW profile \cite{Cirac_Blatt_Zoller_Phillips_1992}. Wavefunctions of highly excited ions outside the LD regime, of practical interest to Doppler cooling, span lengthscales comparable to the wavelength (Fig.~\ref{fig:Dopplerscheme}b), and hence, compromised SW cooling performance is expected as compared to within the LD regime as treated in \cite{Cirac_Blatt_Zoller_Phillips_1992}. To study the cooling behavior beyond the LD regime, here we employ numerical Monte-Carlo simulation of the Lindblad master equation and include higher-order expansion terms (see Appendix). Our simulation quantifies the enhancement of Doppler cooling using SWs as compared to RWs both outside and inside the LD regime, including effects of saturation from sideband excitation. We also discuss experimental configurations of carrier-free Doppler cooling based on integrated photonic addressing \cite{vasquez_control_2023, Mehta_Zhang_Malinowski_Nguyen_Stadler_Home_2020}.

In particular, we simulate the eight-level $\Ca$ structure as an example with a $\hat \pi$-polarized cooling beam ($\Omega_c$, $\Delta_c$) addressing the $\S \leftrightarrow \P$ transition and a repumper ($\Omega_r = 2\pi \times 10$ MHz, $\Delta_r$) with polarization $\hat \epsilon_p = \hat \pi/\sqrt{2} +(\hat{\sigma}_+-\hat{\sigma}_-)/2$ coupling between $\D \leftrightarrow \P$ as shown in Fig.~\ref{fig:Dopplerscheme}a, and with a quantizing magnetic field of $B = 1$~G. Time evolution of the expectation phonon number $\n(t)$ starting from a thermal state $\n(0)$ is extracted from Monte-Carlo simulation of the master equation. The cooling rate at a given $\n$ is defined as $W_c$, and the cooling limit $\nss$ is the final steady-state phonon number.

\begin{figure}
    \centering
    \includegraphics[width=0.49\textwidth]{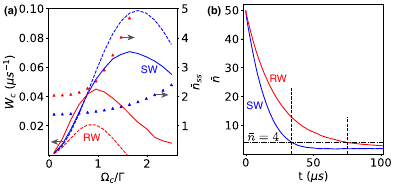}
    \caption{(a) Cooling rate (lines) and limit (points) for $\omega_m = 2\pi \times 3$ MHz and $\Delta_c = -\Gamma/2$. The cooling rate EF is $1.5$ for $\n = 50$ (solid line), and $4.9$ for $\n = 5$ (dotted line). (b) SW and RW cooling trajectory for $\omega_m = 2\pi \times 3$ MHz. Using $\Delta_c= -0.5\Gamma$ and starting at $\n =50$, SW Doppler cooling reaches $\n =4$ within $34 ~\mu s$ with Rabi frequency $\Omega_c = 1.8\Gamma$, while RW Doppler cooling takes $75 ~ \mu s$ with $\Omega_c= 0.8\Gamma$.}
    \label{fig:8LDoppler_sim}
\end{figure}

Fig.~\ref{fig:8LDoppler_sim}a shows $W_c$ in the full eight-level system as a function of $\Omega_c$ for $\omega_m = 2\pi \times 3$ MHz. $\Delta_c$ is set to $-\Gamma_c/2$ which optimizes the cooling limit $\nss$. For small $\Omega_c$, RW and SW Doppler cooling rates increase quadratically with $\Omega_c$ both inside ($\n = 5$) and outside the LD regime ($\n = 50$) before saturation. We define a cooling rate enhancement factor ($\mathrm{EF}$) as a ratio of the maximum achievable cooling rate in either case, i.e. $\mathrm{EF} \equiv W_{SW\max}/W_{RW\max}$. For higher $\n$, this factor decreases but remains greater than 1, with the maximum cooling rate achieved around the same $\Omega_c$ both outside and near the LD regime. Combined with the improvement in $\nss$, this validates the applicability and advantage of SW Doppler cooling even for initially highly excited ions. Fig.~\ref{fig:8LDoppler_sim}b shows the simulation results of a complete Doppler cooling process from $\n = 50$, outside the LD regime, to the Doppler limit, using the optimal parameters for the cooling rate found in Fig.~\ref{fig:8LDoppler_sim}a. We observe that SW Doppler cooling takes less than half of the time compared to RW to reach $\n = 4$, a reasonable starting point for ground-state cooling.
\begin{figure*}[t!]
    \centering
\includegraphics[width=0.99\textwidth]{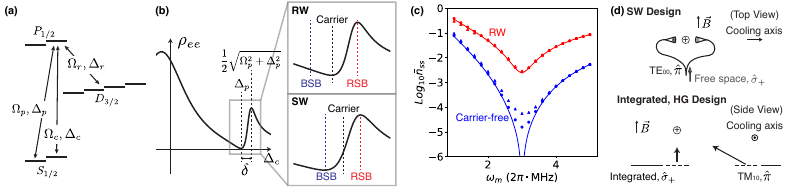}
    \caption{(a) Level structure and fields relevant to EIT cooling of $\Ca$. $\Delta_p$ and $\Delta_c$ are detunings of the pump and cooling laser fields that couple $\ket{S_{1/2}, m_j = -1/2}\leftrightarrow \ket{P_{1/2}, m_j = 1/2}$ and $\ket{S_{1/2}, m_j = 1/2} \leftrightarrow \ket{P_{1/2}, m_j = 1/2}$ respectively. $\Delta_r$ is the repumper's detuning from $D_{3/2}$ and $P_{1/2}$ at $B=0$. (b) Excitation spectrum for a three-level system as a function of the cooling beam's detuning, for fixed $\Delta_p$. Insets show choice of $\Delta_c$ and AC stark shift $\delta$ relative to the motional frequency $\omega_m$ to be cooled, for both RW and SW implementations. (c) Comparison of cooling limit over a range of motional frequencies between simulation results (points) and analytical calculation (solid line) for low-intensity $\Omega_c$ beam. $\Omega_p$ and $\Delta_c$ are optimized for $\omega_{m0} = 2\pi \times 3$ MHz and $\Delta_p = 5 \Gamma$. Triangular and circular dots correspond to RW and SW pump beams, respectively. (d) Experimental schematics of carrier-free EIT cooling using integrated photonics. The first builds on the design for Doppler cooling with a similar integrated SW cooling beam and free-space RW pump beam. The second uses integrated photonics for all beams. A vertically emitted pump beam may achieve high circular polarization purity \cite{Massai_Schatteburg_Home_Mehta_2022} relative to a vertically oriented $B$-field. The cooling beam emitted in TM$_{10}$ has a SW-like beam profile and comes from a shallow angle to satisfy the polarization requirements discussed below. Motional modes with normal mode vector projection along the cooling axis labeled would be cooled in either case.}
    \label{fig:EITscheme}
\end{figure*}

Experimental implementation of carrier-free Doppler cooling may employ the configurations shown in Fig.~\ref{fig:Dopplerscheme}c. Though for practical beam waists, the HG TE$_{10}$'s effective $\eta$ is lower than in a RW, optimal sideband coupling can be recovered by increasing $\Omega_0$ inside the LD regime to hold $\eta \Omega_0$ constant. This allows cooling radial modes of multi-ion strings without requiring precise positioning of multiple ions along a series of SW fringes. Outside the LD regime, however, cooling through higher-order sideband excitations (with coupling strengths proportional to $\eta^k$ for a $k$th order sideband) contribute significantly to cooling, which are attenuated for lower driving beam LD parameter $\eta$ and fixed first-order coupling $\eta \Omega$. In contrast the heating through spontaneous decay on a $k$th order sideband is proportional to $\eta_0^k$ which is independent of the driving beam profile. As a result, the EF in cooling rate achievable with carrier-free cooling decreases outside the LD regime when using lower $\eta$ values. For example, in our eight-level system simulations we observe that the EF for SW cooling with $\eta = \eta_0/4$ compared to the best case RW with $\eta = \eta_0$ drops to nearly unity for $\n = 50$.

Our results quantify the performance of carrier-free Doppler cooling beyond the LD regime previously treated. Though using $\Ca$ in the simulation as an example, similar advantage would hold for other ion species and are more favorable for lower intrinsic $\eta_0$ \cite{xing2023fast}. While the advantage offered is modest in comparison to well within the LD regime, our results indicate that such configurations can still be applied for highly excited ions. Therefore, early cooling stages can share the same cooling beam profile with carrier-free ground-state cooling schemes, for which more substantial enhancements are predicted.

\section{Carrier-free ground-state EIT Cooling}

EIT-based cooling generally requires a pump beam (with Rabi frequency $\Omega_p$) inducing atomic excitation undergoing quantum interference with a weaker cooling beam ($\Omega_c$) in an effective three-level system (Fig.~\ref{fig:EITscheme}a). This results in a Fano profile in the cooling beam's excitation spectrum, with an absorption null at $\Delta_c = \Delta_p$ and a nearby bright peak separated in frequency by the AC Stark shift $\delta  \equiv \frac{1}{2}\sqrt{\Omega_p^2+\Delta_p^2}-\Delta_p$ induced by the pump beam \cite{Morigi_Eschner_Keitel_2000, roos2000experimental}, as shown in Fig.~\ref{fig:EITscheme}b. In the usual configuration with RW optical fields, $\Delta_c$ is tuned to the EIT null to suppress carrier excitation, and $\delta$ is matched to the frequency $\omega_m$ of the motional mode to be cooled, such that its red sideband is preferentially driven (Fig.~\ref{fig:EITscheme}b). The high contrast between the red and blue sideband excitation compared to that achieved with typical dipole transitions employed in Doppler cooling gives rise to cooling with a low final state number \cite{morigi2000ground, roos2000experimental, Lechner_Maier_Hempel_Jurcevic_Lanyon_Monz_Brownnutt_Blatt_Roos_2016}. However, the nonzero blue sideband excitation amplitude limits final phonon number achievable, and furthermore, the Fano profile is usually derived for weak probe beam, failing to accurately describe dynamics for large $\Omega_c$. 

In a carrier-free configuration, the ion can be positioned at a zero-intensity nodal point of the $\Omega_c$ beam. The internal state dressing fundamental to EIT relies on intensity of the pump beam $\Omega_p$, which can be delivered either as a RW or as a SW with anti-node at the ion location, with comparable performance. With the carrier transition nulled by the spatial coherence of the SW $\Omega_c$ field, carrier-free EIT utilizes the pump beam with adjusted parameters to null the blue sideband excitation amplitude. As shown in Fig.~\ref{fig:EITscheme}b, the cooling beam's detuning is chosen such that the blue and red sideband lie at the EIT null and peak, respectively. That is, $\Delta_c = \Delta_p+\omega_{m0}$ and $\delta  = 2\omega_{m0}$. As in resolved sideband cooling, carrier-free EIT in the LD regime then in principle drives strictly the red sideband only, thus allowing extremely low final $\nss$ \cite{Zhang_Wu_Chen_2012}. Additionally, due to suppression of carrier coupling by the cooling beam, higher $\Omega_c$ can be applied to realize faster cooling as compared to the standard RW EIT. Lastly, carrier-free EIT can be expected to offer a larger bandwidth of motional mode frequencies cooled, due to the larger range of detunings from $\Delta_c$ for which high contrast between the red and blue sideband excitation probabilities obtains (see Fig.~\ref{fig:EITscheme}b). Thus as compared to RW-based EIT cooling, carrier-free EIT cooling is expected to enable ground-state cooling of multiple motional modes, simultaneously to lower final phonon numbers. Fig.~\ref{fig:EITscheme}c shows calculated cooling limit in an approximate three-level system for the same pump beam detuning $\Delta_p = 5\Gamma$ and center motional frequency $\omega_{m0} = 2\pi \times 3$~MHz, as a function of motional frequency $\omega_m$. Under the low cooling beam intensity limit, final state phonon number $\nss$ is independent of cooling beam Rabi frequency and whether the pump beam has RW or SW field profile. The agreement between analytic treatment (lines, and see Appendix) and master-equation simulation (points) points to significantly lower $\n_{ss}$ achievable over broader bandwidths using the carrier-free configuration. 

To describe the cooling behavior with large $\Omega_c$ near the saturation limit where the analytical treatment breaks down, we perform master equation simulations of the eight-level electronic system relevant to $\Ca$ at $B = 10$~G and examine the achievable cooling rate, phonon number limit, and bandwidth. Three different couplings are present in the system: a $\hat{\sigma}_+$-polarized pump beam $\Omega_p$ between $\ket{S_{1/2}, m_j = -1/2}$ and $\ket{P_{1/2}, m_j = +1/2}$ with the LD parameter $\eta_p$, a $\hat \pi$-polarized cooling beam $\Omega_c$ between $\ket{S_{1/2}, m_j = +1/2}$ and $\ket{P_{1/2}, m_j = +1/2}$ with the LD parameter $\eta_c$, and a $\hat \epsilon_p$-polarized resonant repumper $\Omega_r$ between the $D_{3/2}$ and $P_{1/2}$, with $\hat \epsilon_p$ defined as in the previous section. Similar to Doppler cooling inside the LD regime, the carrier's relative strength is stronger for smaller $\eta$, suggesting larger possible advantage for low $\eta$. However, carrier-free EIT's EF in cooling rate is not strictly proportional to $\eta_0^{-2}$ because the carrier transition in the RW EIT has also been nulled to first order. Motivated by potential integrated geometries for beam delivery, we also quantify the dependence of these enhancements on imperfect polarization and beam delivery. 

Two practical candidate beam geometries for carrier-free EIT cooling are shown in Fig.~\ref{fig:EITscheme}d, based either on SW or HG delivery of the cooling beam. Motivated by the full level structure relevant to $\Ca{}$ (Fig.~\ref{fig:EITscheme}a),  we consider geometries that best satisfy the polarization requirement of $\Omega_c$ and $\Omega_p$. As detailed in the Appendix, in either case, the motional modes cooled are those with some projection along the cooling beam's field gradient; in contrast to RW-based cooling where cooling rate is proportional to $|\eta_c-\eta_p|^2$, the gradient of the $\Omega_p$ beam does not enter to leading order, and hence both the SW or HG direction can be tuned with respect to the trap axis to cool a combination of axial and radial modes. In the following discussion, we assume the carrier-free cooling beam has spatial gradient along the motional axis such that its LD parameter $\eta_c$ is equal to $\eta_0$. For the RW configuration simulations, to maximize $|\eta_c-\eta_p|$ while accommodating the polarization requirements, we consider RW cooling and pump beam perpendicular to each other and $45^\circ$ from the motional axis such that $\eta_c = -\eta_p = \eta_0/\sqrt{2}$, the choice that maximizes the achievable RW cooling rate.

\begin{figure}[t!]
    \centering
\includegraphics[width=0.49\textwidth]{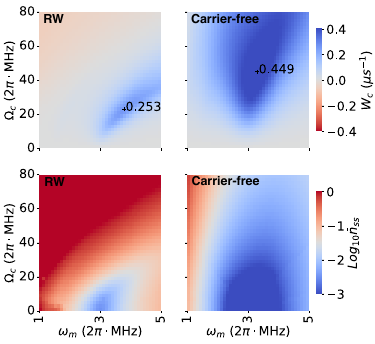}
    \caption{EIT cooling performance as a function $\Omega_c$ and $\omega_m$. The top and bottom rows show the cooling rate and final state phonon number respectively, for typical RW (left) and carrier-free (right) configurations. For all plots, $\Delta_p = 5 \Gamma$, and we choose $\Omega_p = 2 \pi \times 52.2(36.4)$ MHz and $\Delta_c = 5.14 \Gamma(5 \Gamma)$ for SW(RW) EIT as optimal parameters to cool a mode with $\omega_{m0} = 2\pi \times 3$ MHz. Crosses label points of maximal $W_c$. Compared to RW, SW EIT cooling achieves higher cooling rates and lower limits for a larger bandwidth of motional frequency.}
    \label{fig:EITscan}
\end{figure}

 Fig.~\ref{fig:EITscan} shows the simulated $W_c$ and $\n_{ss}$ as a function of the cooling beam's Rabi frequency $\Omega_c$ and different motional frequencies $\omega_m$. The calculations assume a RW pump detuning $\Delta_p = 5\Gamma$ and repumper amplitude $\Omega_r = 2\pi \times 10$ MHz in both cases, with $\Delta_c$ and $\Omega_p$ taken as the optimal values for both RW and SW configurations to target optimal cooling of a motional mode frequency $\omega_{m0} = 2\pi \times 3$ MHz. We find an overall improvement using carrier-free EIT in both cooling rate and final phonon number over a significantly expanded parameter range, cooling $\omega_m/2\pi$ between $1.5$ and $5$ MHz to $\nss<10^{-1}$. 

\begin{figure*}[t!]
    \centering
\includegraphics[width=0.99\textwidth]{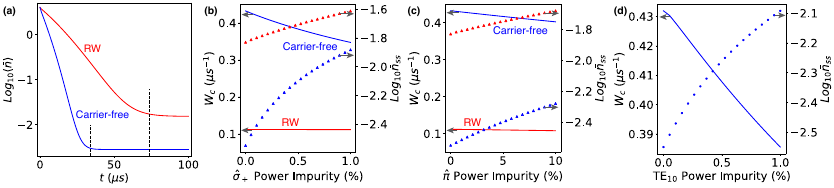}
    \caption{ (a) Cooling trajectory for SW and RW EIT cooling. For a motional mode with frequency $\omega_{m} = 2\pi \times 3$ MHz and starting at $\n =4$, SW EIT cooling reaches $\nss =0.003$ in $34 \,\mu s$, while RW EIT takes $75 \,\mu s$ to achieve $\nss = 0.017$. In both cases, $\Omega_c$ is chosen to maximize $W_c$, corresponding to $\Omega_c = 2\pi \times 42(14)$ MHz for the SW(RW) case. (b) and (c) Cooling rate and limit for parameters chosen to optimally cool $\omega_{m} = 2\pi \times 3$ MHz as a function of the the pump (b) and cooling (c) beam's polarization impurity, assumed evenly distributed in the two undesired polarization components in each case. The results indicate both the sensitivity of the carrier-free cooling's enhancement to the pump beam purity, and the relative insensitivity to the cooling beam impurity. (d) $W_c$ and $\nss$ for a configuration based on first-order HG delivery of the pump beam (Fig.~\ref{fig:EITscheme}d), as a function of the cooling beam's HG TM$_{10}$ mode power impurity (modeled as fraction of power in the fundamental Gaussian instead of the HG$_{10}$ mode).}
    \label{fig:EITsim}
\end{figure*} 

In Fig.~\ref{fig:EITsim}a, we simulate a ground-state cooling instance with the optimal Rabi frequencies that correspond to the highest cooling rate at $\omega_{m0} = 2\pi \times 3$ MHz in Fig.~\ref{fig:EITscan}. Starting near the Dopper cooling limit $\n = 4$, carrier-free EIT cools $\nss$ to $0.003$, $7$ times lower than that of the RW, within half of the time. Furthermore, operation at lower $\Omega_c$ than maximizes $W_c$ would allow substantially lower final $\nss$ (Fig.~\ref{fig:EITscan}a) still. These enhancements in the cooling performance, especially in $\nss$ over a broader motional mode bandwidth than obtained by RW EIT could considerably alleviate the cooling time overhead by reducing or eliminating the need for further sideband cooling.

For integrated implementations enabling practical carrier-free EIT cooling, delivery of the highly pure circular polarization required to realize the effective three-level system presents a challenge \cite{Massai_Schatteburg_Home_Mehta_2022}. To guide design of experimental configurations realizing these functions, we quantify polarization impurities' effects on the predicted performance. We note that implementations at larger magnetic field with larger frequency selectivity can significantly alleviate the required polarization purity \cite{marinelli2020quantum}; however, below we focus on the more challenging polarization purities required for low-field (few G) regimes often used, still assuming $B=10$~G as above. 

Optical access from integrated photonics and $B$-field arrangements impose constraints on design that generally allow practical delivery of one beam with near-perfect polarization purity but compromises in the other's. We consider the effects of both pump and cooling beams' polarization impurities separately. Fig.~\ref{fig:EITsim}b shows cooling rates and limits as a function of the pump beam's power impurity $\epsilon$, evenly distributed between the other two polarization directions such that the polarization vector can be written as $\sqrt{1-\epsilon} \hat{\sigma}_{+} +\sqrt{\epsilon/2} (\hat{\sigma}_{-} +\hat {\pi})$. Fig.~\ref{fig:EITsim}c shows the same for the cooling beam's power impurity. Cooling is clearly sensitive to the strong $\hat\sigma_+$ pump beam's impurity: for impurities larger than approximately 1\% in relative intensity, the advantage of carrier-free EIT cooling is significantly compromised. Cooling performance is significantly more tolerant to the impurity of the $\hat \pi$-polarized cooling beam. 

These polarization and mode purity requirements inform potential experimental configurations for implementation of this cooling scheme. While linear polarization purity can straightforwardly be achieved well below 0.1\% with integrated photonic addressing \cite{mehta2017precise}, achieving $<1\%$ circularly polarization purity presents a challenge for design \cite{he2014chip, spektor2023universal}. Fig.~\ref{fig:EITscheme}d gives two possible experimental schemes, both utilizing integrated delivery of the cooling beam but based either on free-space or integrated delivery of the $\hat \sigma_+$ beam. In the latter case, a vertical B-field is chosen given the likely advantage of a surface-normal-propagating $\hat \sigma_+$ beam with high purity from passively robust structures \cite{Massai_Schatteburg_Home_Mehta_2022, natarajan2024arrayed}; $\pi$-polarized cooling light then corresponds to a vertically polarized  $E$-field, which is approximately achieved via grating emission of a TM-polarized waveguide mode at a shallow emission angle propagating close to parallel with the trap chip. The ${\sim}10\%$ tolerance on the $\hat \pi$-beam polarization purity allows for emission angles for quasi-TM-polarized cooling beams of approximately $72.5 ^{\circ}$ from normal.

Naturally, the advantage of carrier-free EIT cooling relies on the stable driving of the ions at the nodal points of the cooling beam. For the three-level system, ref. \cite{Zhang_Wu_Chen_2012} shows that $\nss$ degrades by $3 \times$ for a positional deviation from the cooling beam's node of $0.017\lambda$ with $\lambda$ the cooling beam's wavelength. Consistent with results from our eight-level simulations (see Appendix), this corresponds to a position deviation from the SW node of order 10 nm, to a location with approximately $1\%$ of the SW's anti-node intensity. Ion addressing in a phase-stable SW with a similar level of position accuracy has been demonstrated with integrated photonics \cite{vasquez_control_2023}, indicating that the required positioning precision is practically achievable. The bottom design in Fig.~\ref{fig:EITscheme}d utilizes fully integrated beams \cite{Massai_Schatteburg_Home_Mehta_2022}, with cooling field delivered through TM$_{10}$ mode. Fig.~\ref{fig:EITsim}d shows that simulated $\nss$ in the eight-level system degrades by an amount consistent with above for 1\% power impurity in the fundamental mode, agreeing with \cite{Zhang_Wu_Chen_2012}. This sensitivity sets a design goal for experiments to suppress the fundamental mode's relative intensity to approximately 1\% level to avoid unwanted scattering. 

Our results indicate significant possible gains in ground-state cooling via carrier-free EIT cooling, simultaneously in motional mode frequency bandwidth, final phonon number, and cooling rate. By potentially significantly reducing requirements for subsequent sideband cooling, these techniques may alleviate a key bottleneck in trapped-ion system runtime. Polarization purity in pump beam delivery is quantified as a key challenge for implementation at low magnetic fields, along with mode purity requirements when utilizing HG mode drives. 

Integration of photonic elements for the UV wavelengths required for $\Ca$ laser cooling presents a challenge but may leverage alumina or hafnia/alumina waveguides recently demonstrated for low-loss blue and UV photonics \cite{west2019low, Sorace-Agaskar_Kharas_Yegnanarayanan_Maxson_West_Loh_Bramhavar_Ram_Chiaverini_Sage_et, Jaramillo2024} and also employed in recent integrated trapped-ion experiments \cite{kwon2024multi, clements2024sub}. Similar schemes can be implemented at longer wavelengths utilizing e.g. barium ions; and extension of schemes utilization carrier suppression for EIT cooling to ion species with nonzero nuclear spin, including with alternate beam and polarization configurations, is an interesting direction for further exploration \cite{huang2024electromagnetically}. Experiments towards realization of the proposed schemes with $\Ca{}$ using foundry-fabricated trap devices incorporating both alumina and silicon nitride waveguides \cite{Beck_Home_Mehta_2024} are in progress.

\section{Conclusion}
By selectively driving desired sideband transitions, carrier-free cooling at nodal points of SW or HG modes allows advantages over cooling with running waves in cooling speed and final phonon number limit. Modest advantages obtain for Doppler cooling of highly excited ions from well beyond the LD regime, with significant enhancements predicted for ground-state cooling via EIT including in motional mode frequency bandwidth addressed. The proposed schemes utilize either simple SW or HG modes that can be stably and robustly delivered to ions in integrated photonic architectures. Furthermore, carrier nulling may also allow resolved sideband cooling at high rates that with RW fields would be limited by carrier excitation due to power broadening. Our results indicate potential for scalable photonics to assist in alleviating key physical limitations of trapped-ion systems, and inform experimental realizations in progress.

\section{Acknowledgments} 
We thank Erich Mueller for helpful discussions. We acknowledge support from an NSF CAREER award (2338897), IARPA via ARO through Grant Number W911NF-23-S-0004 under the ELQ program, the Alfred P. Sloan Foundation, and Cornell University.

\section{Data availability} 
The data that support the findings of this article are openly available \cite{PQEGroup/LC_sim_2025}.
\appendix

\section{Lindblad master equation beyond the LD regime}

The dynamics of a trapped ion in a harmonic potential, interacting with a laser field, can be described using the Lindblad master equation. Inside LD regime, Eq.~(\ref{eq:one}) is approximated to first order $\langle e| V_{dip} |g\rangle \approx \frac{\Omega_0}{2} (1+i\eta (\hat{a}+\hat{a}^\dag))$, resulting in a carrier and two first-order sideband transitions. The strongest present coupling, carrier transition, governs the internal states of the ion. This allows us to approximate the system by decoupling the external motional states from the internal dynamics and describe the cooling behavior with the rate equation \cite{Cirac_Blatt_Zoller_Phillips_1992, Stenholm_1986}. These approximations fail when going outside the LD regime or when sideband transitions become the leading order coupling as in the carrier-free cooling configurations considered here. Thus, to investigate the dynamics outside the LD regime and quantify the enhancement limit of carrier-free cooling, we directly simulate the Lindblad master equation with higher-order terms using a Monte-Carlo method.

Higher-order terms in the dipole interaction Hamiltonian come immediately from Taylor expansions as in Eq.~(\ref{eq:one}); higher-order dissipative terms can be derived similarly with some subtleties. For a two-level system with isotropic spontaneous decay rate $\Gamma$, the Liouvillian 
\begin{eqnarray}
\L^d \rho &=& \frac{1}{2} \Gamma  \int_{-1}^1 dx \cdot \frac{1}{2} \big( 2\hat{A} \rho \hat{A}^\dag -  \hat{A} \hat{A}^\dag \rho  - \rho \hat{A} \hat{A}^\dag\big) \nonumber\\
&=& \frac{1}{2} \sum_m[2\hat{c}_m \rho \hat{c}_m^\dag -  \hat{c}_{m} \hat{c}_m^\dag \rho  - \rho \hat{c}_m \hat{c}_m^\dag]
\label{eq:two}
\end{eqnarray}
with $\hat{A} = e^{ixk_0 \hat{R}} \hat{\sigma}^-, \hat{\sigma}^- = |g\rangle \langle e|$, $x = \cos(\theta)$ for emission angle $\theta$ with respect to motional axis \cite{Cirac_Blatt_Zoller_Phillips_1992}.
Using the Baker–Campbell–Hausdorff formula, $e^{i x k_0 \hat{R}} = e^{i x \eta_0 (\hat{a}^{\dag} + \hat{a})}= e^{-\eta_0^2 x^2 /2} e^{i x \eta_0 \hat{a}^{\dag}} e^{i x \eta_0 \hat{a}}$. Denote $\hat{b}_k = \frac{(i \eta_0 \hat{a}^{\dag})^k}{k!}, \hat{d}_k = \frac{(i \eta_0 \hat{a})^k}{k!}$ ($\hat{b}_j=\hat{d}_j = 0 \text{ if } j<0 \text{ or } j>n$), and Taylor expanding $\hat{A}$ to the $n$th term, we can then calculate the jump operator in the Liouvillian with $\pm m$ ($0 \leq m\leq n$) change in motional states as 
\begin{widetext}
\begin{equation}
\hat{c}_m = \sqrt{\Gamma} \hat{\sigma}^- \otimes \sum^{n}_{i=0} \left( \sqrt{\int_{-1}^1 dx \cdot \frac{1}{2} e^{-\eta_0^2 x^2} x^{4i+2m}} \hat{b}_i \hat{d}_{i+m} + \sqrt{\int_{-1}^1 dx \cdot \frac{1}{2} e^{-\eta_0^2 x^2} x^{4i-2m}}\hat{b}_i \hat{d}_{i-m} \right).
 \label{eq:higherdecay}
\end{equation}
\end{widetext}

\section{Analytical EIT treatment}
\begin{figure}[t!]
    \centering
\includegraphics[width=0.45\textwidth]{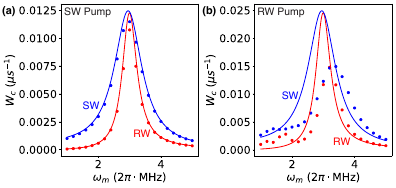}
    \caption{Comparison of simulation (points) and analytical results (lines) for cooling rate over $\omega_m$ for the low-intensity limit of cooling beam. The pump beam's intensity and cooling beam's detuning are optimized for $\omega_{m0} = 2\pi \times 3$ MHz and $\Delta_p = 5 \Gamma$. Blue and red represent results for cooling beam delivered as a SW and RW, respectively. (a) uses a SW pump beam with the ion at an antinode, with $\eta_c =\eta_0$ and $\Omega_c = 2\pi \times 2$ MHz for both SW and RW cooling beam. (b) uses RW pump beam. We choose $\eta_c = -\eta_p  = \eta_0/\sqrt{2}$ for RW EIT and $\eta_c = \eta_0$ for the SW cooling beam as discussed in the previous section. The same LD parameters are used in the simulations shown in previous sections. Lastly, we choose $\Omega_c = 2\pi \times 2 \sqrt{2}$ MHz for SW cooling beam and $\Omega_c = 2\pi \times 2$ MHz for RW cooling beam so that their cooling rates at $\omega_{m0}$ agree in quantum regression theorem calculation. The presence of pump-beam sideband couplings present in (b) distorts the good agreement between the analytical and master-equation treatment in (a); both treatments show the enhancement in motional frequency bandwidth associated with SW cooling beam delivery. }
    \label{fig:QRT}
\end{figure}
Our analytical description of carrier-free EIT cooling follows the approach of \cite{Zhang_Wu_Chen_2012} for an ideal three-level system. A three-level system with pump beam of carrier strength $\Omega_{p0}$ and sideband strength $\eta_p\Omega_{p1}$ (coupling states $\ket {r}$ and $\ket e$) and cooling beam of carrier strength $\Omega_{c0}$ and sideband strength $\eta_c\Omega_{c1}$ (coupling states $\ket g$ and $\ket e$) can be represented by a rotating-frame Hamiltonian: 
\begin{eqnarray}
\hat H_0/\hbar 
    = && \nu a^{\dag} a -\Delta_c |e\rangle \langle e| - (\Delta_c - \Delta_p) |r\rangle \langle r| \nonumber \\&& + \frac{1}{2}\big(\Omega_{c0}\hat{\sigma}_{gx} +\eta_c\Omega_{c1}\hat{\sigma}_{gy}\hat{R}+\Omega_{p0}\hat{\sigma}_{rx} +\eta_p\Omega_{p1}\hat{\sigma}_{ry} \hat{R}\big)\nonumber \\ 
    \equiv &&  \nu a^{\dag} a + \hat H_\mathrm{int} - \hat F \hat R
\end{eqnarray}
with $\nu$ the motional frequency. We define the force operator $\hat{F} \equiv -\frac{\Omega_{c1}}{2}\eta_c\hat{\sigma}_{gy}-\frac{\Omega_{p1}}{2}\eta_p\hat{\sigma}_{ry}$, where $\hat{\sigma}_{gy} \equiv -i(\ket{e}\bra{g}-\ket{g}\bra{e})$ and $\hat{\sigma}_{ry} \equiv -i(\ket{e}\bra{r}-\ket{r}\bra{e})$. The calculation proceeds by approximating the internal state density matrix as the steady-state solution of the optical bloch equations under the action of the internal-state Hamiltonian $\hat H_\mathrm{int}$, and then considering the effect of the relatively weak forcing terms as a perturbation. 

The fluctuation spectrum corresponding to the forcing term is
\begin{eqnarray}
    \mathcal{S}(\nu) = &\lim_{t\rightarrow \infty}\int_0^{\infty} d\tau e^{i\nu \tau} \langle \hat F(\tau + t) \hat F(t) \rangle \nonumber \\
    = &\lim_{t\rightarrow \infty}\{(\frac{\Omega_{c1}}{2} \eta_c)^2 \mathcal{F} [\langle \hat{\sigma}_{gy}(\tau + t) \hat{\sigma}_{gy}(t)] \rangle \nonumber \\
    &+ (\frac{\Omega_{p1}}{2} \eta_p)^2\mathcal{F} [\langle \hat{\sigma}_{ry}(\tau + t) \hat{\sigma}_{ry}(t)]\nonumber\\
    &+\eta_c\eta_p\frac{\Omega_{c1} \Omega_{p1}}{4} \big(\mathcal{F} [\langle \hat{\sigma}_{gy}(\tau + t) \hat{\sigma}_{ry}(t) \rangle] \nonumber \\
    &+ \mathcal{F} [\langle \hat{\sigma}_{ry}(\tau + t) \hat{\sigma}_{gy}(t) \rangle] \big) \},
    \label{eq:QRTeq}
\end{eqnarray}
where $\mathcal{F}$ denotes the Fourier transform with respect to $\tau$. Using the quantum regression theorem \cite{Gardiner_Zoller_2004}, we can solve for an analytical expression of (\ref{eq:QRTeq}) from the steady-state solution to $\hat H_\text{int}$.

When the ion sits at the nodal point, we have $\Omega_{c0} = 0$ in the $\hat H_\text{int}$. Thus, only the first term in (\ref{eq:QRTeq}) is nonzero. Taking the leading order of $\Omega_{c1}$,
\begin{eqnarray}
    \text{Re} [\mathcal{S}(\nu)] = \frac{2\eta_c^2\Omega_{c1}^2\Gamma (\Delta +\nu )^2}{4\Gamma^2 (\Delta +\nu )^2+\big(\Omega_{p0}^2-4(\Delta +\nu )(\Delta_c +\nu) \big)^2},
    \nonumber
\end{eqnarray}
where $\Gamma = \Gamma_c + \Gamma_p$ and $\Delta = \Delta_c-\Delta_p$. When $\Delta = \nu$, this gives the cooling rate in \cite{Zhang_Wu_Chen_2012} for ideal carrier-free EIT, which does not depend on $\eta_p$. 

For the standard RW EIT cooling with $\Omega_{c0} = \Omega_{c1}$ and $\Omega_{p0} = \Omega_{p1}$, taking the leading order of $\Omega_c$ again, (\ref{eq:QRTeq}) reduces to a simple expression when $\Delta = \Delta_c-\Delta_p$, proportional to $(\eta_c - \eta_p)^2 \Omega_{c1}^2$. 

The analytical treatment above assumes that sideband couplings do not affect the stead-state solution to the optical Bloch equation, valid in the LD regime and in the low intensity limit for both $\Omega_{c1}$ and $\Omega_{p1}$. In reality, the effect of the pump beam's sideband coupling on the internal dynamics is always non-negligible. Fig.~\ref{fig:QRT} shows cooling rates for EIT using SW and RW pump beam, comparing the simulation with the analytical results from (\ref{eq:QRTeq}) under the low-intensity limit of cooling beam. For the SW pump beam, we consider an ion positioned at the SW's antinode where only the carrier is present. The discrepancy between the simulation and analytical results for RW pump beam indicates the need for full master equation simulation.

\section{Position Deviation}
\begin{figure}[t!]
    \centering
\includegraphics[width=0.45\textwidth]{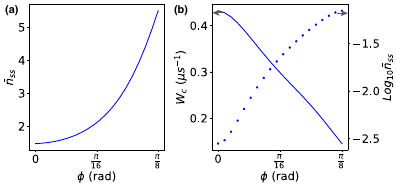}
    \caption{Sensitivity of both SW Doppler and EIT cooling to positional offset from the SW node, for $\omega_m = 2\pi \times 3$ MHz. Positional offset is defined in terms of phase offset $\phi$ within the SW field with spatial dependence $\sin (\bm{k}\cdot\hat{\bm{R}}+ \phi)$. (a) Doppler cooling limit with $\Delta_c= -0.5\Gamma$ and $\Omega_c = 1.8\Gamma$ as in Fig.~\ref{fig:8LDoppler_sim}b. (b) EIT cooling rate and limit with optimal parameters for $\omega_m = 2\pi \times 3$ MHz, $\Delta_p = 5\Gamma$ and $\Omega_c = 2\pi \times 42$ MHz as in Fig.~\ref{fig:EITsim}a. }
    \label{fig:Deviation}
\end{figure}

The carrier-free cooling schemes depend strongly on the suppression of carrier transition by placing the ion at the intensity null. Thus, the effect of ion's position inside the SW is discussed previously for both Doppler and EIT cooling \cite{Cirac_Blatt_Zoller_Phillips_1992, Zhang_Wu_Chen_2012}. To complete the discussion, we simulate the cooling dynamics for the eight-level $\Ca$ at different $\phi$ inside the SW $\sin (\bm{k}\cdot\hat{\bm{R}}+ \phi)$ in Fig.~\ref{fig:Deviation}. The results suggest that when the ion is positioned within $|\phi|\lesssim\frac{\pi}{16}$, the final EIT cooling limit remains within one order of magnitude of that at the SW node. For Doppler cooling, the same positional tolerance results in negligible impact on final phonon number. For purely counterpropagating SW beams, this corresponds to a positional deviation of $|\Delta x| \lesssim  \frac{\lambda}{32}$, i.e. a positional accuracy of within approximately $\pm 10$ nm. Previous observations of trapped-ion control in integrated photonic devices indicate the practicality of such positioning \cite{vasquez_control_2023}. We note further that for SW beams intersecting at an angle (i.e. with a lower wavevector projection along the motional mode axis) the positional accuracy required to achieve $\phi \lesssim \frac{\pi}{16}$ is further relaxed. 
 
\section{Motional bandwidth for carrier-free Doppler cooling}
\begin{figure}[]
    \centering
\includegraphics[width=0.45\textwidth]{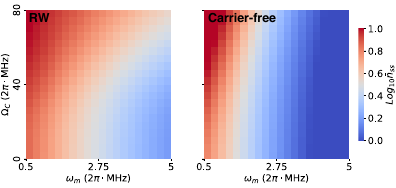}
    \caption{Doppler cooling limit in RW (left) and SW (right) fields as a function of $\Omega_c$ and $\omega_m$ with $\Delta_c = -\Gamma/2$, calculated from the eight-level simulation.}
    \label{fig:Dopplermotional}
\end{figure}

In Fig.~\ref{fig:EITscan}, we observe that carrier-free EIT when optimized for a center motional frequency can cool a larger motional bandwidth as compared to RW EIT. For Doppler cooling, both RW and carrier-free cooling should be able to cool a broad range of motional frequencies. Within the LD regime and for low cooling beam amplitude $\Omega_c$ set to $\Delta_c = -\Gamma/2$, we expect $\nss = \frac{\Gamma}{2\omega_m}$ for RW and $\nss = \frac{\Gamma}{4\omega_m}$ for carrier-free Doppler cooling \cite{Cirac_Blatt_Zoller_Phillips_1992}. This behavior is apparent in the simulation results for $\bar n_\mathrm{ss}$ from the eight-level simulation, particularly at large $\Omega_m$ for which the LD limit applies (Fig.~\ref{fig:Dopplermotional}).

\bibliography{reference}

\end{document}